\documentclass[aps,pra,reprint,superscriptaddress,longbibliography]{revtex4-1}

\usepackage[english]{babel}
\usepackage{amsmath}    % need for subequations
\usepackage{graphicx}   % need for figures
\usepackage[lofdepth,lotdepth, caption=false]{subfig}
\usepackage{verbatim}   % useful for program listings
\usepackage{color}      % use if color is used in text
\usepackage{hyperref}   % use for hypertext links, including those to external documents and URLs
\usepackage{dcolumn}

\begin{document}
\newcommand{\degrees}{$^\circ$C}

\title{Dependence of superconductivity in Cu$_{x}$Bi$_{2}$Se$_{3}$ on quenching conditions}

\author{J.~A.~Schneeloch}
\email{jschneeloch@bnl.gov}
\affiliation{Condensed Matter Physics and Materials Science Department, Brookhaven National Laboratory, Upton, New York 11973, USA}
\affiliation{Department of Physics and Astronomy, Stony Brook University, Stony Brook, NY 11794, USA}

\author{R.~D.~Zhong}
\affiliation{Condensed Matter Physics and Materials Science Department, Brookhaven National Laboratory, Upton, New York 11973, USA}
\affiliation{Materials Science and Engineering Department, Stony Brook University, Stony Brook, NY 11794, USA}

\author{Z.~J.~Xu}
\altaffiliation[Present address: ]{Lawrence Berkeley National Lab, 1 Cyclotron Road, Berkeley, CA 94720}
\affiliation{Condensed Matter Physics and Materials Science Department, Brookhaven National Laboratory, Upton, New York 11973, USA}

\author{G.~D.~Gu}
\affiliation{Condensed Matter Physics and Materials Science Department, Brookhaven National Laboratory, Upton, New York 11973, USA}

\author{J.~M.~Tranquada}
\affiliation{Condensed Matter Physics and Materials Science Department, Brookhaven National Laboratory, Upton, New York 11973, USA}

%\date{\today}

\begin{abstract}

Topological superconductivity, implying gapless protected surface states, has recently been proposed to exist in the compound Cu$_{x}$Bi$_{2}$Se$_{3}$. Unfortunately, low diamagnetic shielding fractions and considerable inhomogeneity have been reported in this compound.  In an attempt to understand and improve on the finite superconducting volume fractions, we have investigated the effects of various growth and post-annealing conditions. With a melt-growth (MG) method, diamagnetic shielding fractions of up to 56\% in Cu$_{0.3}$Bi$_{2}$Se$_{3}$ have been obtained, the highest value reported for this method. We investigate the efficacy of various quenching and annealing conditions, finding that quenching from temperatures above $560^{\circ}$C is essential for superconductivity, whereas quenching from lower temperatures or not quenching at all is detrimental. A modified floating zone (FZ) method yielded large single crystals but little superconductivity. Even after annealing and quenching, FZ-grown samples had much less chance of being superconducting than MG-grown samples. From the low shielding fractions in FZ-grown samples and the quenching dependence, we suggest that a metastable secondary phase having a small volume fraction in most of the samples may be responsible for the superconductivity.  

\end{abstract}

\maketitle

\section{Introduction}

Materials with topologically nontrivial electronic structures have gained a great deal of attention recently, both for their unique physics and for their potential applications. Whereas conventional insulators or superconductors can be adiabatically transformed into topologically trivial states \cite{ando_topological_2015, hasan_colloquium:_2010}, symmetry-protected topological insulators (TI) and superconductors (TSC) \cite{roy_topological_2008, schnyder_classification_2008, kitaev_periodic_2009, qi_time-reversal-invariant_2009, qi_topological_2011} cannot be so transformed without breaking certain symmetries, resulting in surface states robust to many kinds of perturbations. For example, the surfaces of time-reversal invariant TSCs would host Majorana fermions, which are potentially useful for the low decoherence needed for quantum computation and are also desirable for investigating their unique properties, such as being their own antiparticles \cite{fu_superconducting_2008, nayak_non-abelian_2008, leijnse_introduction_2012}.

The recently discovered \cite{hor_superconductivity_2010} compound Cu$_{x}$Bi$_{2}$Se$_{3}$, in which Cu is intercalated between layers of the TI Bi$_{2}$Se$_{3}$, was soon proposed to be a TSC \cite{fu_odd-parity_2010}. Many experiments have probed topological properties in this material, with mixed findings. The strongest evidence thus far for TSC has come from point-contact spectroscopy data \cite{sasaki_topological_2011, kirzhner_point-contact_2012, chen_point-contact_2012} showing zero-bias conductance peaks (ZBCPs) that may be indicative of unconventional superconductivity.  Calculations showed that, for a two-orbital model of the band structure, every possible unconventional pairing symmetry should be topologically nontrivial \cite{fu_odd-parity_2010, sasaki_topological_2011}, suggesting that Cu$_{x}$Bi$_{2}$Se$_{3}$ could be a TSC. On the other hand, scanning tunneling spectroscopy measurements \cite{levy_experimental_2013} showed no such peaks except when superconductor-insulator-superconductor junctions were accidentally formed, suggesting that Cu$_{x}$Bi$_{2}$Se$_{3}$ is instead an s-wave superconductor and highlighting the difficulties in making point-contact measurements on Cu$_{x}$Bi$_{2}$Se$_{3}$. In addition, spectroscopy with normal-metal/superconductor junctions \cite{peng_absence_2013} showed ZBCPs for a transparent barrier but not for a finite barrier, raising more doubts about unconventional superconductivity in Cu$_{x}$Bi$_{2}$Se$_{3}$. On the other hand, one theoretical study \cite{mizushima_dirac-fermion-induced_2014} suggested that the link between unconventional superconductivity and a ZBCP is not as simple as earlier suggested, and that an odd-parity superconductor with a cylindrical Fermi surface could account for a peak absence. Findings from ARPES experiments have also been mixed, with the finding of a conical dispersion relation for Cu$_{0.12}$Bi$_{2}$Se$_{3}$\cite{wray_observation_2010, wray_spin-orbital_2011, wang_structural_2011, tanaka_evolution_2012, kondo_anomalous_2013, lahoud_evolution_2013} and other characteristics of the band structure seen as favoring TSC, but a later study showing that the Fermi surface encloses an even number of time-reversal invariant momentum points casting doubt on TSC\cite{lahoud_evolution_2013}. In addition, specific heat data \cite{kriener_bulk_2011}, nuclear magnetic resonance data \cite{fu_odd-parity_2014}, and what have been claimed as anomalously high superconducting transition temperatures ($T_{c}\sim3.5$~K) for the measured carrier concentrations ($\sim10^{-20}$~cm$^{-3}$) \cite{hor_superconductivity_2010, kriener_bulk_2011, ando_topological_2015} have been interpreted as favoring TSC, though the specific heat behavior may have a more conventional explanation \cite{sandilands_doping-dependent_2014}.

Unfortunately, the large inhomogeneity and low diamagnetic shielding fractions of crystals have contributed to the difficulty in studying this compound, and it would be highly desirable to improve crystal quality and obtain clearer results. Since the discoverers of Cu$_{x}$Bi$_{2}$Se$_{3}$ reported only a 20\% shielding fraction at their lowest achievable temperature \cite{hor_superconductivity_2010} and did not show zero resistance, there have been doubts as to the bulk nature of the superconductivity  \cite{kriener_bulk_2011}. They used a melt-growth method, where one seals Cu, Bi, and Se in a quartz ampoule, heats to the melting point, slowly cools the mixture as crystallization occurs, then quenches. Other groups \cite{das_spin-triplet_2011, kirzhner_point-contact_2012, bay_superconductivity_2012} used similar methods, obtaining similar shielding fractions, and two of these groups \cite{kirzhner_point-contact_2012, bay_superconductivity_2012} were able to measure zero resistance in their samples. More recently, a superconducting fraction of 35\% was reported for the melt-growth method \cite{lawson_quantum_2012}. A different method involving quenching from the liquid state while using precursor ingredients to avoid Cu$_{2}$Se production \cite{kondo_reproducible_2013} has also been reported to improve shielding fraction. Alternatively, an electrochemical method has been used \cite{kriener_electrochemical_2011, ando_topological_2015} to intercalate Cu in pre-grown Bi$_2$Se$_3$ crystals, with reported zero resistance,  shielding fractions of up to $\sim$70\%, and specific heat data indicating bulk superconductivity. 

In this paper, we report the effects of various growth conditions on diamagnetic shielding fraction in Cu$_{x}$Bi$_{2}$Se$_{3}$, highlighting changes in conditions that result in qualitative increases in the likelihood of superconductivity.  We have found shielding fractions as high as 56\%, showing that it is possible to obtain substantial shielding fractions using the melt-growth method; however, while the occurrence of superconductivity is generally reproducible, the magnitude of the shielding fraction is not. We investigated the effects of various quenching and annealing conditions, highlighting the importance of quenching, which has been investigated in detail for the electrochemical \cite{kriener_electrochemical_2011} and quench-from-liquid \cite{kondo_reproducible_2013} methods and prescribed but not studied in detail for the melt-growth method \cite{hor_superconductivity_2010}. While not quenching results in little superconductivity, its effects appear to be reversible by subsequent annealing and quenching. Annealing at temperatures of 560\degrees\ or higher before quenching was found to be essential for superconductivity, whereas somewhat lower temperatures were actually detrimental to superconductivity. Large single crystals of Cu$_{x}$Bi$_{2}$Se$_{3}$ grown by the floating zone method are generally non-superconducting. These observations suggest that the phase responsible for the superconductivity is metastable.

The rest of the paper is organized as follows.  In the next section, we describe the methods we used for sample growth and treatment, as well as the characterization methods.  In Sec.~III, we present our results.  Their significance is discussed in Sec.~IV, where we point out similarities with K$_x$Fe$_{2-y}$Se$_2$, another system for which quenching is essential for achieving superconductivity.  A brief conclusion is given in Sec.~V.

\section{Materials and methods}

We used two growth methods, the melt-growth (MG) method and a modified floating-zone growth (FZ) method.  For the MG method \cite{hor_superconductivity_2010}, stoichiometric amounts of Cu, Bi, and Se were sealed in an ampoule that was sealed within another ampoule (double-sealed) under 0.2 bar Ar, then the ampoule was placed horizontally in a small box furnace, heated well above the melting point to 840\degrees, jostled to mix the ingredients, cooled to 640\degrees \ at $18$\degrees/h, and then quenched in liquid nitrogen. The inner ampoules were 2mm thick, had 10 mm inner diameter, and were $\sim$15 cm long; the outer ampoules had similar dimensions and were 1mm thick. Typically, Cu$_{x}$Bi$_{2}$Se$_{3}$ was made this way in 50g batches. For the FZ method, stoichiometric amounts of Cu, Bi, and Se were vertically premelted in an ampoule (sealed similarly to the MG method), then zone-melted from bottom to top in-ampoule using an image furnace at a rate of 0.6 mm/h, resulting in single crystals of up to several cm in length. For samples grown by both the MG and FZ methods, the impact of post-annealing (hereafter, ``annealing'') was explored, where pieces were sealed in a single ampoule (typically, 1mm thick, 10mm inner diameter, and $\sim$15 cm long) under 0.2 bar Ar, heated at 580\degrees \ for 4 h, then quenched in liquid nitrogen. For all growth methods, the Cu, Bi, and Se were of purities 99.999\%, 99.999\%, and 99.995\%, respectively, and used without pretreatment. 

There have been a number of reports on the best choice of Cu concentration $x$ for superconductivity, with superconductivity found within $0.1 \leq x \leq 0.3$ for the MG method \cite{hor_superconductivity_2010}, $0.1 \leq x \leq 0.6$ for the electrochemical method \cite{kriener_electrochemical_2011}, and $0.03 \leq x \leq 0.5$ for a quench-from-liquid method \cite{kondo_reproducible_2013}. Our preliminary data suggested $0.25 \leq x \leq 0.35$ as an optimal range, so we have focused our efforts on $x=0.3$ when comparing different growth and post-annealing conditions.

For magnetic measurements, we used a Quantum Design Magnetic Properties Measurement System, with the field $H$ applied within the cleavage plane ($H \perp c$) to minimize the demagnetization effect. To mitigate the effect of magnetic relaxation \cite{das_spin-triplet_2011}, we waited 30 seconds after applying the field before measuring the magnetic response. To calculate shielding fraction, we took data under zero-field cooling at $T=1.7$ K for at least two different fields within $0 < H \leq 2$ Oe, and used the slope of the line fit through these points, the sample's mass, and the density to calculate the shielding fraction. The density was determined from a cylindrical piece of an FZ ingot of nominal Cu$_{0.35}$Bi$_{2}$Se$_{3}$ composition. For the resistance measurements, the four-probe method was used with current applied in the $ab$-plane. The temperature was controlled by the magnetometer cryostat. We performed annealing on pieces thicker than $1$ mm and then cleaved samples from the inside of these pieces for measurement. A direct comparison of the superconducting properties of cleaved samples before and after annealing was not possible since annealing thin ($< 1$ mm) samples invariably resulted in loss of superconductivity. We suspect this loss was due to preferential evaporation of Se, which was proposed to be responsible for the n-type doping of Bi$_{2}$Se$_{3}$ \cite{hor_p-type_2009, xia_observation_2009}. To characterize the composition of our materials, we used a JEOL 7600F scanning electron microscope (SEM) equipped with energy-dispersive x-ray (EDX) analysis capabilities located at the Center for Functional Nanomaterials at Brookhaven National Laboratory.   The compositions measured as a function of position are reported in Fig.~\ref{fig:ResistivityMagnetization}; in contrast, the values of Cu concentration cited in the text are generally the nominal concentrations, corresponding to the elemental mixtures from which samples were grown.

\section{Results}
\begin{figure}[t]
\begin{center}
\includegraphics[width=8.6cm]
{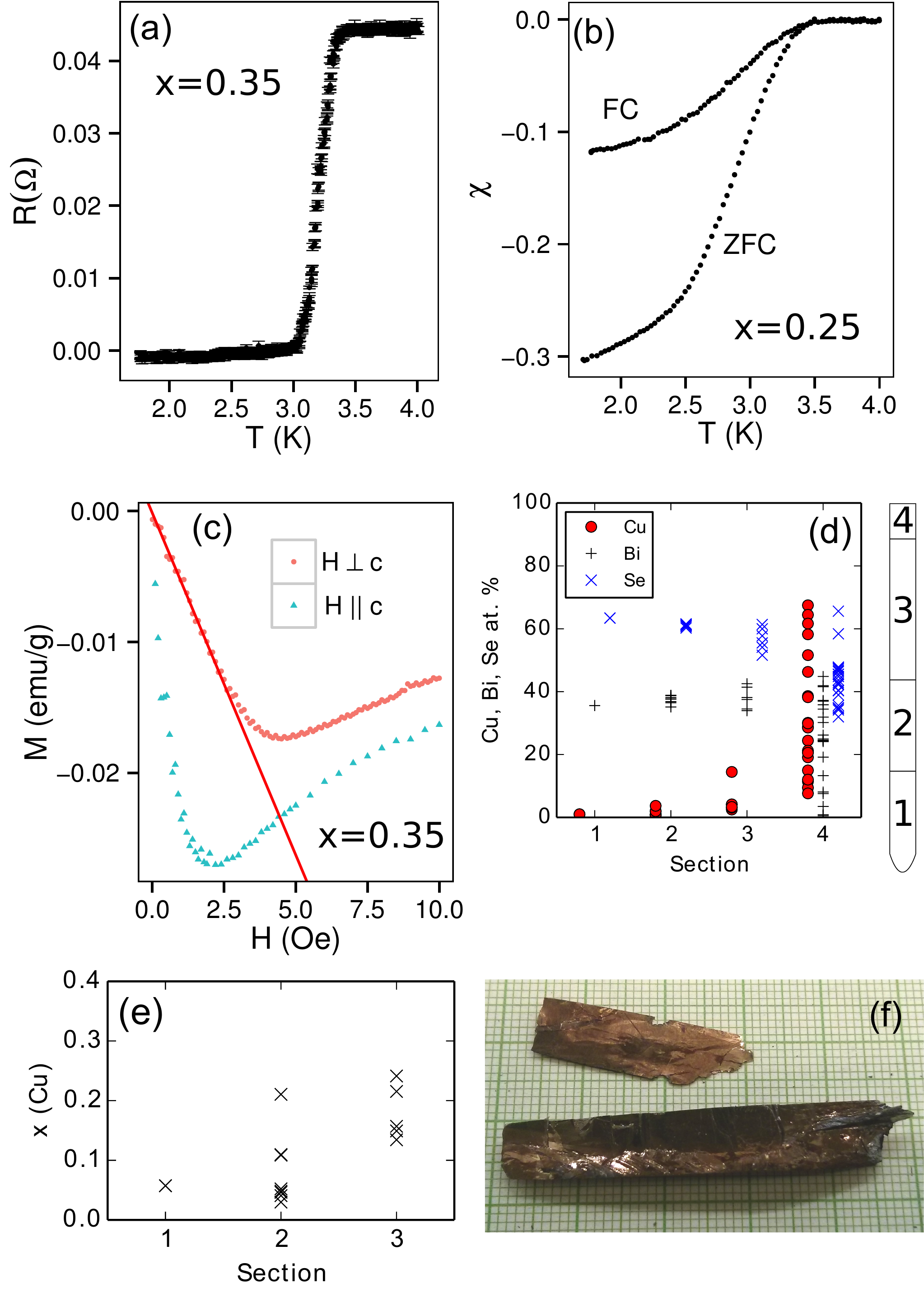}
\caption{\label{fig:ResistivityMagnetization} (Color online) (a) Resistance vs.\ temperature for a single crystal of $x=0.35$. (b) Magnetic susceptibility $\chi$ vs.\ temperature for a single crystal with $x=0.25$, 2.2 mg, and $\sim$30\% shielding fraction at 1.7 K. The applied field was 2 Oe with $H \perp c$, with both ZFC and field cooling (FC) shown. (c) Magnetization curve for a single crystal with $x=0.35$ near 1.7 K, for both $H \perp c$ and $H \parallel c$, under zero field cooling (ZFC); the shielding fraction was found to be 48\% at 1.7 K. The best-fit line used for shielding fraction calculation has been plotted. In (b) and (c), two measurements were taken at each field and temperature respectively, then averaged and plotted. (d,e) Composition as determined by EDX measurements for samples taken from different sections of an $x=0.25$ FZ ingot. The sections are ordered by increasing height. (d) Atomic percents of Cu, Bi, and Se. (e) $x$ for the ratio Cu$_{x}$:Bi$_{2}$ for the first three sections. (f) Photo of crystals cleaved from section 3 of the same ingot. The smallest division of the grid is 1 mm.}
\end{center}
\end{figure}

In Fig.~\ref{fig:ResistivityMagnetization}(a) the resistance parallel to the \textit{ab}-plane is plotted as a function of temperature for a sample with $x=0.35$. Each plotted resistance point corresponds to the average of ten consecutive measurements; error bars show standard deviation. The resistance drops to nearly zero from roughly 3.4 to 3.0 K, indicating superconductivity, with a small further decrease down to 1.7 K. Magnetoresistance measurements (not shown) indicated that $B_{c2}$ is between 1 and 7 T, consistent with previous reports \cite{hor_superconductivity_2010, kriener_bulk_2011, levy_experimental_2013}. 

In Fig.\ \ref{fig:ResistivityMagnetization}(b), we plot susceptibility as a function of temperature for a 2.2 mg sample with $x=0.25$ having $T_c \approx 3.5$ K and a shielding fraction of $\sim$30\% at 1.7 K. The applied field was $H = 2$ Oe with $H \perp c$. The magnetization has a broad transition and seems likely to continue dropping well below $\sim$1.7 K, our lowest achievable temperature. We have seen similarly broad curves for all of our other magnetization vs.\ temperature data, as well as for data reported elsewhere \cite{hor_superconductivity_2010, kriener_bulk_2011, das_spin-triplet_2011, das_direct_2013, lawson_quantum_2012}. Generally, measured T$_{c}$ values varied within $2.5 \leq T \leq 3.6$, consistent with the $T_{c}$ range reported for the electrochemical method\cite{kriener_electrochemical_2011}, but unlike in ref.\ \cite{kriener_electrochemical_2011} there is no apparent correlation between T$_{c}$ and nominal $x$ value in our data. This lack of correlation may be due to the increased inhomogeneity expected for the MG method.

Magnetization data as a function of field are plotted in Fig.\ \ref{fig:ResistivityMagnetization}(c) for a 5.1 mg sample with $x=0.35$ and $\sim$48\% shielding fraction at 1.7 K. Both $H \perp c$ and $H \parallel c$ are plotted. The highest shielding fraction that we measured was $56$\% for another $x=0.35$ sample with similar growth and annealing conditions. The demagnetization effect was neglected since it is expected to be small for field applied to a flat crystal within its plane. The dimensions for the 48\% and 56\% shielding fraction samples were roughly $1.7 \times 2.1 \times 0.25$ mm$^3$ and $1.4 \times 0.8 \times 0.2$ mm$^3$, respectively. These samples were brittle and their size was small due to attempting to isolate portions with the highest shielding fractions; typical dimensions for other samples were the same thickness or less, and longer. These shielding fractions are higher than any previously reported shielding fractions for the MG method \cite{hor_superconductivity_2010, das_spin-triplet_2011, kirzhner_point-contact_2012, bay_superconductivity_2012, lawson_quantum_2012, kondo_reproducible_2013} and comparable to those of the electrochemical method \cite{kriener_bulk_2011}. Our magnetization curves for $H \perp c$ and $H \parallel c$ are similar to magnetization curves previously reported\cite{kriener_bulk_2011, das_spin-triplet_2011}, with $H_{c1}$ significantly larger for $H \perp c$ than for $H \parallel c$. We note that the minima of the magnetization curves for $H \perp c$ varied from roughly 4 Oe to 10 Oe for different crystals, even for the same nominal composition, growth, and annealing conditions, and there was no apparent correlation with shielding fraction.

To get insight into how the Cu was incorporated into Bi$_{2}$Se$_{3}$, we performed EDX measurements on samples from one of our $x=0.25$ FZ ingots having good crystal quality, as shown in Figures \ref{fig:ResistivityMagnetization}(d) and \ref{fig:ResistivityMagnetization}(e). We note that there was a decrease of the measured Cu to Bi ratio with increasing electron accelerating voltage (and little change for the Se to Bi ratio), possibly indicating a strong depth dependence in Cu composition, so a high voltage of 25 keV was used to probe the bulk composition to a greater extent. The ingot was divided into four sections, ordered by height (see Fig.\ \ref{fig:ResistivityMagnetization}(d), right); sections 1 to 4 had lengths of 3, 2.5, 5.5, and 1 cm, respectively. In Fig.~\ref{fig:ResistivityMagnetization}(d), atomic percentages for Cu, Bi, and Se are plotted for pieces selected from each section. Fig.\ \ref{fig:ResistivityMagnetization}(e) shows a portion of the same data, but in terms of the $x$ in the ratio Cu$_{x}$:Bi$_{2}$. Unfortunately, EDX-measured $x$ values varied widely for the same nominal $x$. Even MG Cu$_{0.05}$Bi$_{2}$Se$_{3}$ samples had similar EDX $x$ values (not plotted) as for nominal $x=0.25$ or $0.3$ samples. This variation is evident in Fig.\ \ref{fig:ResistivityMagnetization}(e). The reason for the inconsistent EDX results is uncertain, though it is possible that Cu tends to segregate in certain planes, making those planes easiest to cleave and thus most likely to be probed, or that Cu segregates to the surface soon after cleaving. Nevertheless, in Fig.\ \ref{fig:ResistivityMagnetization}(d) we see clear evidence of Cu-rich, non-Bi$_{2}$Se$_{3}$ related compositions in section 4, with a large increase in Cu fraction and a deviation from a Bi:Se ratio of 2:3. These data suggest that, for most of the ingot, Cu$_{x}$Bi$_{2}$Se$_{3}$ of lower-than-nominal Cu concentration crystallized as the liquid zone passed through the starting materials, making the liquid zone increasingly Cu-rich until the Cu precipitated out at the top of the ingot.

In Fig.\ \ref{fig:ResistivityMagnetization}(f), we show a photo of a crystal cleaved from the inside of the FZ ingot whose EDX data is plotted in Fig.\ \ref{fig:ResistivityMagnetization}(d) and \ref{fig:ResistivityMagnetization}(e); we see that crystals many cm long can be cleaved from the inside of FZ ingots, unlike for the MG method where such large crystals are only found at the top surface (corresponding to the free surface of the melt). A copper coloring which usually appears on Cu$_{x}$Bi$_{2}$Se$_{3}$ samples after some time has passed \cite{hor_superconductivity_2010} is present. Usually, this coloring appeared slowly, on the order of weeks or months, but we have seen similar coloring already present on crystals taken from an ingot that was quenched from a liquid melt. Despite appearance changes, we have observed that crystals initially found to be superconducting remain superconducting for many months afterward, but there is variation between samples and some loss has been observed. For example, the sample initially measured to have 48\% shielding fraction was measured to have 36\% shielding fraction 19 months later, whereas two other samples, each with 7\% to 8\% shielding fractions, still had shielding fractions within this range 7 months later.

\begin{figure}[h]
\begin{center}
\includegraphics[width=8.6cm]
{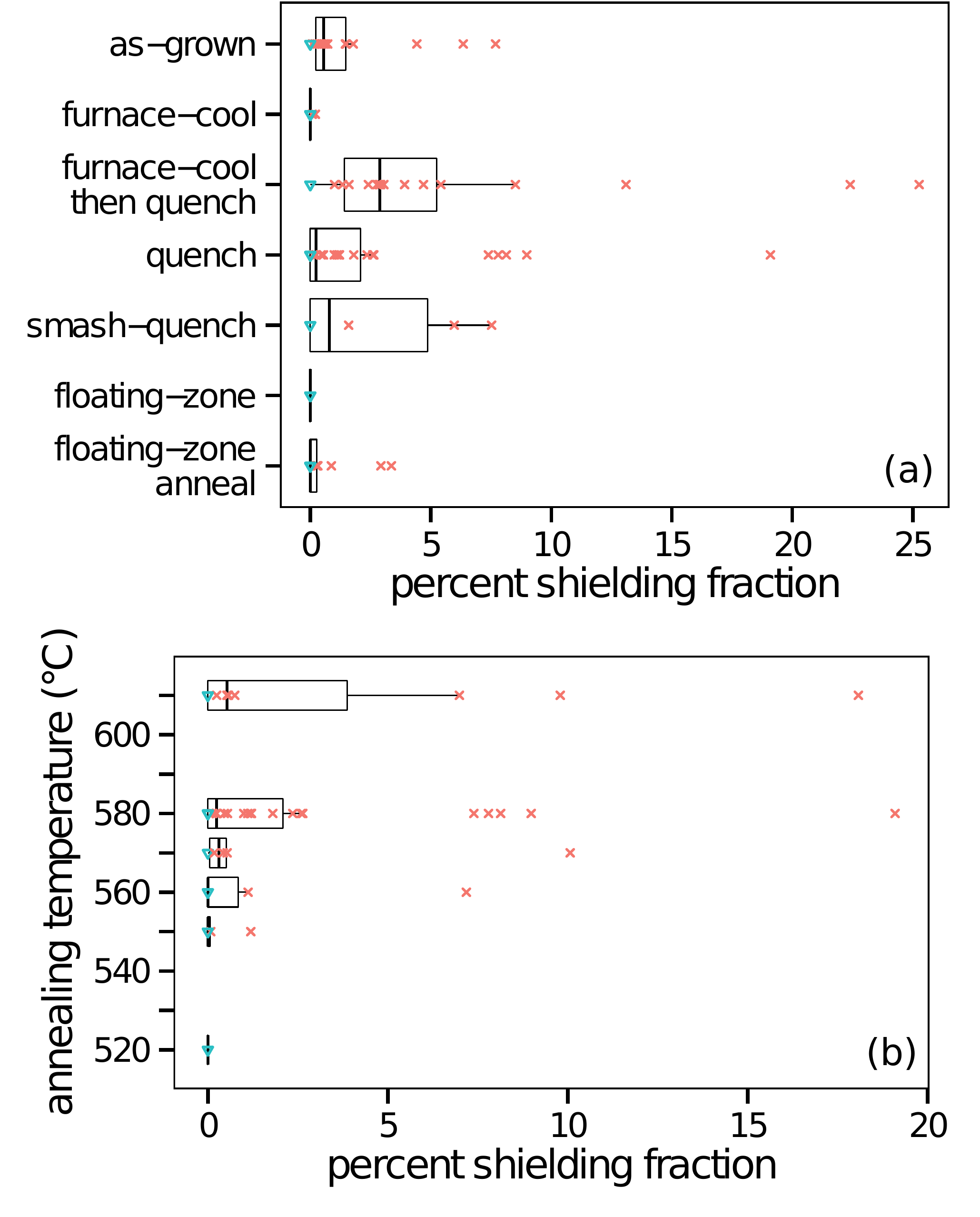}
\caption{\label{fig:shieldingFractions} (Color online) (a) Shielding fractions obtained for samples of Cu$_{0.3}$Bi$_{2}$Se$_{3}$ obtained for various growth and annealing conditions. See text for description of conditions. The numbers of samples plotted in each row from top to bottom are 19, 12, 18, 31, 6, 6, and 18. (b) Shielding fractions for MG Cu$_{0.3}$Bi$_{2}$Se$_{3}$ samples obtained after annealing and quenching at the displayed temperatures. The numbers of samples plotted in each row from top to bottom are 11, 31, 6, 6, 6, and 6. In (a) and (b), for each distribution of shielding fractions (including non-superconducting samples), the boxplot indicates the median, quartiles, and the most extreme values not greater than 1.5 times the interquartile range from the median; the triangles represent samples with negligible diamagnetic response.}
\end{center}
\end{figure}

Shielding fractions for Cu$_{0.3}$Bi$_{2}$Se$_{3}$ crystals subjected to various growth and annealing conditions are plotted in Fig.\ \ref{fig:shieldingFractions}(a). After growth (which included quenching), crystals were cleaved from the inside of the ingots and measured (``as-grown''), with remaining parts of the ingots being annealed and exposed to different cooling conditions: quenching in liquid nitrogen and cooling within a minute (``quench''); smash-quenching, where the ampoule is smashed open in liquid nitrogen and cooled within seconds (``smash-quench''); or furnace-cooling, in which the furnace is shut off and the ampoule is cooled within several hours (``furnace-cool''). Some of the samples that were furnace-cooled were later annealed and quenched (``furnace-cool then quench''). In addition, as-grown and annealed FZ samples (``floating-zone'' and ``floating-zone anneal'') were measured. Due to the large variation in shielding fractions even for the same conditions, we focus on cases where there is a qualitative difference between the shielding fraction distributions for different conditions. 

A few findings are readily apparent in Fig.\ \ref{fig:shieldingFractions}(a). First, annealing followed by furnace-cooling was clearly detrimental to superconductivity, with most of the samples measured having no diamagnetic response, whereas annealing followed by quenching had results comparable to the as-grown samples (which were also quenched), showing the importance of quenching for superconductivity in Cu$_{x}$Bi$_{2}$Se$_{3}$. Pieces that were annealed and furnace-cooled and then annealed and quenched showed relatively large shielding fractions, suggesting that the detrimental effects of annealing and furnace-cooling appear to be reversible by subsequent annealing and quenching. We see no clear difference between the quenching and smash-quenching data, suggesting that cooling to room temperature within a minute is sufficiently fast to allow superconductivity. 

The FZ method was much less effective than the MG method for making superconducting samples. None of the six as-grown FZ samples included in Fig.~\ref{fig:shieldingFractions}(a) were superconducting, and we have also found that magnetic responses at $H=10$ Oe for crystals from other FZ ingots with $x=0.25$ or $0.35$ (not plotted) were also very low.  By annealing and quenching, it is possible to induce superconductivity; however, the shielding fraction tends to be low compared to the MG samples. It should be noted that the samples measured tended to be good single crystals cleaved from the middle of the FZ ingot.  

In Fig.~\ref{fig:shieldingFractions}(b) we show the dependence of shielding fraction on annealing temperature for MG Cu$_{0.3}$Bi$_{2}$Se$_{3}$ samples that were quenched after annealing. (All pieces were originally quenched from 580\degrees\ or higher after growth.) We see that large shielding fractions were achievable after quenching from an annealing temperature above 560\degrees, with lower temperatures appearing to result in little superconductivity relative to either the as-grown or quenched-above-560\degrees\ sample sets. The temperature-dependence agrees with data reported for the electrochemical method \cite{kriener_electrochemical_2011}, with superconductivity found after annealing between 530\degrees\ and 620\degrees, whereas no superconductivity was found between 500\degrees\ to 520\degrees.

\section{Discussion}
There are a number of aspects of the crystal growth and superconductivity in Cu$_{x}$Bi$_{2}$Se$_{3}$ which need explanation. Why does the superconductivity vary so wildly from sample to sample, with most samples having low shielding fraction but a few having $>$$50$\% shielding fraction? Why is quenching necessary? Why do FZ-grown crystals have low shielding fractions? 

We can address some of these questions by comparing the MG and FZ methods. The MG method should result in more inhomogeneity before quenching on a macroscopic scale than the FZ method. The large Cu concentration near the top of the FZ ingot in Fig.~\ref{fig:ResistivityMagnetization}(d) and the relatively low Cu concentrations obtained elsewhere suggest that for slow cooling the solubility of Cu in Cu$_{x}$Bi$_{2}$Se$_{3}$ is low. Combined with the decreasing melting point of Cu$_{x}$Bi$_{2}$Se$_{3}$ with increasing Cu concentration \cite{babanly_phase_2010, kriener_electrochemical_2011}, it seems that cooling slowly from the liquid phase should result in inhomogeneity, with lower-Cu Cu$_{x}$Bi$_{2}$Se$_{3}$ crystals solidifying before higher-Cu compositions, as suggested by others\cite{kondo_reproducible_2013}. In the FZ method, the Cu-rich regions would be deposited toward the end of the ingot under ideal crystal growth conditions, but in the MG method these regions would be deposited in the interior of the ingot, increasing the inhomogeneity. If impurity phases were responsible for the superconductivity in Cu$_{x}$Bi$_{2}$Se$_{3}$, this would provide a natural explanation for the relative lack of superconductivity in FZ-grown crystals, though another possibility is that the Cu concentration in the FZ-grown crystals is too low. 

For more insight into the source of the inhomogeneity and the superconductivity in Cu$_{x}$Bi$_{2}$Se$_{3}$, we consider two cases: primary-phase, where the superconductivity arises from regions having the same phase as Bi$_{2}$Se$_{3}$; or secondary-phase, where the superconductivity arises from an impurity phase. In both of these cases, quenching above 560\degrees\ is presumed to preserve the configuration at high temperatures, while annealing and quenching within 550\degrees\ to 520\degrees\ (and probably lower) is presumed to equilibrate the system to the lower-temperature configuration. 

For the primary-phase case, the superconductivity should arise from sufficiently highly-intercalated Bi$_{2}$Se$_{3}$ regions, since our lower-than-nominal Cu-intercalated Bi$_{2}$Se$_{3}$ FZ crystals were non-superconducting. (It should be noted, though, that a Cu$_{x}$Bi$_{2}$Se$_{3}$ thin film with $x=0.12$ was found non-superconducting\cite{shirasawa_structure_2014} even with a carrier concentration higher than reported for bulk superconducting samples \cite{hor_superconductivity_2010}.) To account for the necessity of quenching, we presume that highly-intercalated Bi$_{2}$Se$_{3}$ may be stable at high temperatures but may decompose to less-intercalated Bi$_{2}$Se$_{3}$ and Cu-rich impurity phases upon slow cooling, a decomposition which should be reversible with further annealing and quenching. While this scenario is consistent with some observations, we feel it does not provide a natural explanation for the inhomogeneity; assuming that the copper is intercalated homogeneously above 560\degrees, one should only expect inhomogeneity after quenching in this scenario if only a small portion of the sample cooled quickly enough. 

For the secondary-phase case, the role of quenching would be to preserve a phase separation rather than a single phase. Specifically, at high temperatures we assume a phase separation between a less Cu-intercalated Bi$_{2}$Se$_{3}$ phase and a Cu-rich impurity phase, with the impurity phase being only metastable at lower temperatures. The Cu-rich phase would be the most obviously responsible for the superconductivity considering the low shielding fractions and the majority Bi$_{2}$Se$_{3}$ phase present in most samples. 

This situation is similar to that of K$_{x}$Fe$_{2-y}$Se$_{2}$, another layered superconducting system that must be quenched from high temperature in order to obtain superconductivity \cite{guo10}. While the dominant phase, with ordered Fe vacancies, exhibits antiferromagnetic order with large magnetic moments \cite{ye11}, it is now clear that the superconductivity in K$_{x}$Fe$_{2-y}$Se$_{2}$ is driven by a second, epitaxial phase that may be vacancy free \cite{cai12,yuan12,ricc11,li12}. The second phase can only be obtained by quenching; slow cooling yields only the vacancy-ordered antiferromagnet \cite{han12}. $T_{c}$ values remain nearly constant around 30~K with varying composition \cite{yan_electronic_2012}, suggesting that the secondary phase maintains a similar composition despite changes in nominal composition. Though $T_{c}$ varies to a greater extent in Cu$_{x}$Bi$_{2}$Se$_{3}$, it also appears to be bounded, with the lowest reported $T_{c}$ being 2.2 K and shielding fractions being maximized for intermediate $T_{c}$ in electrochemically-doped samples \cite{kriener_electrochemical_2011}. Given the limited shielding fraction in Cu$_{x}$Bi$_{2}$Se$_{3}$ and the similarities to K$_{x}$Fe$_{2-y}$Se$_{2}$ discussed above, as well as the lack of superconductivity in the FZ-grown crystals, which are expected to have higher crystal quality than MG-grown crystals, the possibility that the superconductivity might be due to a yet-to-be-identified second phase deserves consideration.

\section{Conclusions}
In summary, we have investigated various growth and annealing conditions on Cu$_{x}$Bi$_{2}$Se$_{3}$ and identified their effects on shielding fractions. Shielding fractions as high as 56\% have been measured, showing that a substantial shielding fraction is possible for the MG method, though typical values are much lower. Quenching after annealing at a sufficiently high temperature was shown to be crucial to obtaining superconductivity, as equilibrated samples from furnace-cooling are generally non-superconducting. One can recover superconductivity in such samples by subsequent annealing and quenching. FZ-grown crystals had negligible superconductivity. The anneal and quench treatment can yield significant shielding fractions in MG samples, but for the large FZ-grown crystals, the fraction is always small.  EDX measurements on the FZ ingots indicate that the Cu concentration is substantially lower than the nominal value in the feed rod throughout most of the growth process.   

The fact that quenching is essential indicates that the superconducting phase is metastable.  In combination with the observation that shielding fractions are usually very small, we suggest that the superconductivity might be driven by a secondary phase that makes up a small volume fraction of the sample.

\section*{Acknowledgments}
This work was supported as part of the Center for Emergent Superconductivity, an Energy Frontier Research Center funded by the U.S. Department of Energy, Office of Science, Office of Basic Energy Science, and benefitted from facilities at the Center for Functional Nanomaterials, which is funded by the same Office. The work was performed at Brookhaven, which is funded through Contract No.~DE-SC00112704.

\end{document}